\documentclass[referee]{raa_rb}
\usepackage[pagebackref=true]{hyperref}

\usepackage{graphicx,times}
\usepackage{natbib}
\usepackage{amssymb,amsmath}
\bibpunct{(}{)}{;}{a}{}{,}
\usepackage{xcolor}
\bibpunct{(}{)}{;}{a}{}{,}

\definecolor{dkblue}{RGB}{54, 86, 169}

\newcommand{\EQ}[1] {Equation~(\ref{#1})}
\newcommand{\SEC}[1] {Section~\ref{#1}}
\newcommand{\APP}[1] {Appendix~\ref{#1}}
\newcommand{\FIG}[1] {Figure~\ref{#1}}

\newcommand{\VEC}[1] {{\boldsymbol{{ #1}}}}

\begin{document}
\title{Searching for the nano-Hertz stochastic gravitational wave background 
	 with the Chinese Pulsar Timing Array Data Release I}

 \volnopage{ {\bf 20XX} Vol.\ {\bf X} No. {\bf XX}, 000--000}
   \setcounter{page}{1}

 \author{Heng Xu
	 \inst{1}, Siyuan Chen\inst{2,3}, Yanjun Guo\inst{4},
	 Jinchen Jiang\inst{1}, Bojun Wang\inst{1}, Jiangwei Xu\inst{3,2,1},
	 Zihan Xue\inst{3,2,1}, R. Nicolas Caballero\inst{5,2}, Jianping Yuan\inst{6}, Yonghua Xu\inst{7}, Jingbo Wang\inst{8}, Longfei Hao \inst{7}, Jingtao Luo \inst{9}, Kejia Lee \inst{*3,1,2,10,11}, 
	Jinlin Han\inst{1}, Peng Jiang\inst{1}, Zhiqiang Shen\inst{12}, 
	 Min Wang\inst{6}, Na Wang\inst{5}, Renxin Xu\inst{2,3,13},
	  Xiangping Wu\inst{1},
	 Richard Manchester\inst{14},  
	 Lei Qian\inst{1}, Xin Guan\inst{1}, Menglin Huang\inst{1},
	 Chun Sun\inst{1}, Yan Zhu\inst{1} }
	 \institute{National Astronomical Observatory, Chinse Academy of Sciences, 
	 Beijing 100101, P.~R.~China;\\
	 \and
	  Kavli Institute for Astronomy and Astrophysics, Peking University, 
	 Beijing 100871, P.~R.~China\\
	 \and
	 Department of Astronomy, School of Physics, Peking University, Beijing 100871, P.~R.~China; {\it 
	 kjlee@pku.edu.cn}\\
\and
Max-Planck institut für Radioastronomie, Auf Dem Hügel, Bonn, 53121, Germany
	 \and
	Hellenic Open University, School of Science and Technology, 26335 Patras, Greece
	 \and
	 Xinjiang Astronomical Observatory, Chinese Academy of Sciences, Urumqi 830011, Xinjiang, P.~R.~China
	 \and
	  Yunnan Astronomical Observatories, Chinese Academy of Sciences, Kunming 650216, Yunnan, P.~R.~China
	  \and
	   Institute of Optoelectronic Technology, Lishui University, Lishui, Zhejiang, 323000, P.~R.~China
	   \and
	   	National Time Service Center, Chinese Academy Of Sciences, Xi'an 710600, P.~R.~China
	 \and
  Key Laboratory of Radio Astronomy and Technology, Chinese Academy of Sciences,  Beijing, 100101, P.~R.~China
  \and
	 Beijing Laser Acceleration Innovation Center, Huairou, Beijing, 101400, P.~R.~China	 
	 \and
	 Shanghai Astronomical Observatory, Chinese Academy of Sciences, Shanaghai 200030, P.~R.~China	
	  \and
	 State Key Laboratory of Nuclear Physics and Technology, School of Physics, Peking University, Beijing 100871, P.~R.~China
	 \and
	  Australia Telescope National Facility, CSIRO Space and Astronomy,  P.O. Box 76, Epping NSW 1710, Australia
	 \\
	 {\small Received 20XX Month Day; accepted 20XX Month Day}
}

\abstract{Observing and timing a group of millisecond pulsars (MSPs) with high rotational stability enables the direct detection of gravitational waves (GWs). The GW signals can be identified from the spatial correlations encoded in the times-of-arrival of widely spaced pulsar-pairs. The Chinese Pulsar Timing Array (CPTA) is a collaboration aiming at the direct GW detection with observations carried out using Chinese radio telescopes. This short article serves as a `table of contents' for a forthcoming series of papers related to the CPTA Data Release 1 (CPTA DR1) which uses observations from the Five-hundred-meter Aperture Spherical radio Telescope (FAST). Here, after summarizing the time span and accuracy of CPTA DR1, we report the key results of our statistical inference finding a correlated signal with amplitude $\log A_{\rm c}= -14.4 \,^{+1.0}_{-2.8}$ for spectral index in the range of $\alpha\in [-1.8, 1.5]$ assuming a GW background (GWB) induced quadrupolar correlation. The search for the Hellings-Downs (HD) correlation curve is also presented, where some evidence for the HD correlation has been found that a 4.6-$\sigma$ statistical significance is achieved using the discrete frequency method around the frequency of 14 nHz. 
We expect that the future International Pulsar Timing Array data analysis and the next CPTA data release will be more sensitive to the nHz GWB, which could verify the current results.
\keywords{(stars:) pulsars: general --- gravitational waves --- methods: 
statistical --- methods: observational}
}

\authorrunning{H. Xu et al. }            
\titlerunning{CPTA DR1, Searching for nHz stochastic GW background}  
   \maketitle

\section{Introduction}
\label{sect:intro}

A Pulsar Timing Array \citep[PTA; ][]{fb1990} is an array of pulsars, which are regularly observed. The times-of-arrival (TOAs) are measured for pulses that we see beams of electromagnetic waves emitted by the pulsars sweeping over the Earth. As the directions of the radiation beam and the pulsar rotational axis do not coincide, we observe this radiation as regular pulses synchronized to the pulsar rotation \citep{gold69}. By extracting correlated signatures in pulsar TOAs, it is possible to detect GWs \citep{HD83,JHLM05}, measure masses \citep{CHM10,cgl+2018} and orbital elements of Solar-system planets \citep{GLLC19}, and study the stability of international \citep{HCM12,HGC20} and local \citep{llc+2020} atomic time standards. 

At present, the PTA experiment is the only known effective method to detect GWs in the nanohertz (nHz) band.  It was predicted that the orbiting and merger of supermassive black-hole binaries (SMBHBs) would create a stochastic nHz GW background \citep{SESANA13}, while alternative channels, including cosmic strings \citep{kib1976,SBS12} and relic GWs from early universe processes \citep{gri2005,2013PhRvD..87l4012Z,LMS16}, are also possible.  All these sources provide the GWB targets for PTA experiments.
While in addition, PTAs are also able to detect GWs from single SMBHB systems \citep[e.g.][]{jll+2004,svv2009,lwk+2011}.
At present, six major regional PTAs exist: \emph{Parkes Pulsar Timing Array} (PPTA, \citealt{PPTA0, PPTA,2022ApJ...932L..22G}), \emph{European Pulsar Timing Array} (EPTA, \citealt{EPTA,2022MNRAS.509.5538C}), \emph{North American Nanohertz Observatory for Gravitational Waves} (NANOGrav, \citealt{NANOGRAV09, NANOGRAV}), \emph{Indian Pulsar Timing Array} (InPTA, \citealt{InPTA}), \emph{South Africa Pulsar Timing Array} (SAPTA, \citealt{2022PASA...39...27S}), and \emph{Chinese Pulsar Timing Array} \citep{CPTA}. The \emph{International Pulsar Timing Array}  (IPTA,\citealt{IPTA}) is a consortium of the regional PTA consortia aiming at a better GW sensitivity. Currently, the PPTA, EPTA, NANOGrav, and InPTA are formal IPTA members, while SAPTA and CPTA are IPTA observers. 

Here, we report the progress of CPTA efforts. The current status of CPTA observations and data quality are summarized in \SEC{sec:obs}. Our statistical inference for the amplitude and spectral index of a nHz stochastic GWB are presented in \SEC{sec:gwa}. \SEC{sec:hd} includes the search for the GW-induced quadrupolar correlation (i.e. the HD curve). Discussion and conclusions are in \SEC{sec:con}.

\section{Observations and data}
\label{sec:obs}

The CPTA DR1 consists of TOA measurements and pulsar timing models for 57 MSPs (Xu et al., in prep.). The data covers the time span between April 2019 and September 2022 . Observations were conducted using the Five-hundred-meter Aperture Spherical radio Telescope \citep[FAST; ][]{Jiang19SCPMA}, in Guizhou province, China. For most MSPs, the majority of the observations were taken with a cadence of approximately two weeks. Pulsars with smaller timing errors, e.g., PSR J1713$+$0747, were observed more frequently since February 2020 under a CPTA extension proposal. We excluded data of PSR J1713$+$0747 after MJD=59319 in our analysis because of the abrupt profile change event \citep{2021ATel14642....1X,2021MNRAS.507L..57S,2021RNAAS...5..167L,2022arXiv221012266J}. The observations were carried out with the central beam of the 19-beam receiver \citep{Jiang20RAA} within the frequency range from 1.0 to 1.5 GHz.

Search mode data were recorded with a \textsc{ROACH 2}-based system\footnote{Reconfigurable Open Architecture Computing Hardware: \url{http://casper.berkeley.edu/wiki/ROACH2}}. The data were folded offline in 30-second intervals using the software package \textsc{DSPSR}~\citep{Straten2011}. With the software package \textsc{PSRCHIVE}~\citep{psrchive}, the data were cleaned by removing radio interference, polarization calibrated (Jiang et al. in prep.), and integrated over frequency and time. The final number of frequency channels was 64 ($\sim 7.8$~MHz resolution), except for PSRs J0218$+$4232 and J0636$+$5129, which have many observations, where the number of frequency channels was 16 in order to reduce the size of the dataset. 
The final integration time for integrated pulse profiles was typically 20 minutes, or shorter than 2.5\% of the binary period. Our TOAs were generated by cross-matching observed pulse profiles with the standard total-intensity templates. Sub-integrations with low signal-to-noise (S/N<8) were removed (less than 5.2\% in total). More details on the dataset can be found in Xu et al. (in prep).

\begin{figure} \centering
	 \includegraphics[width=14.0cm, angle=0]{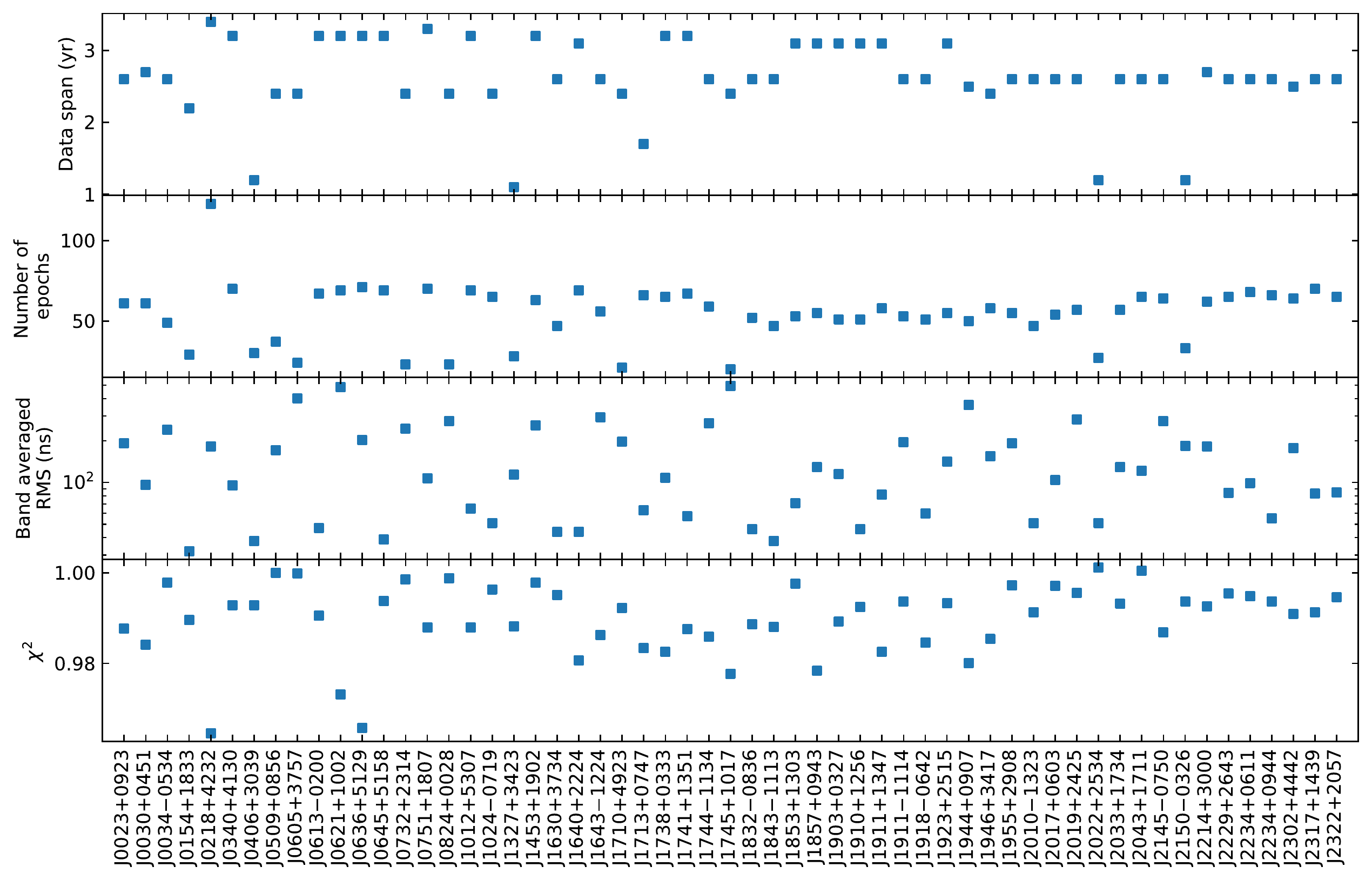}	\caption{Summary of the CPTA 
	 DR1 data quality. From top to bottom, the panels show data span, number of 
	 epochs, weighted root-mean-square (RMS) of frequency integrated residuals, 
	 and reduced $\chi^2$ of timing residual, respectively.  \label{fig:data}}
\end{figure}

The pulsar timing models were created using the software package \textsc{TEMPO2} \citep{tempo22006}. Our noise model for a single pulsar consists of three components: white, red, and dispersion measure (DM) noise. The white noise is characterized using EFAC, EQUAD, and ECORR parameters. EFAC re-scales the TOA measurements error to account for inaccuracies in the process of TOA extraction using the template-matching method \citep{tay1992}, EQUAD adds white noise in quadrature \citep{vlj+2011}, and ECORR \citep{ng2015} models the correlated white noise (phase jitter) among different frequencies in the same epoch (Wang, et al., in prep.). Both red and DM noises are modeled as stochastic stationary processes, assuming power-law spectra, characterized by amplitude and spectral index. Our inference of pulsar noise model parameters is conducted within a Bayesian framework. The definitions of model parameters are well described in the literature (e.g. \citealt{LBJ14,temponest}), of which the conventions are followed in our analysis. Four independent noise analysis software pipelines were used, namely, \textsc{TEMPONEST}\citep{temponest}, \textsc{ENTERPRISE}~\footnote{\url{https://github.com/nanograv/enterprise}}, \textsc{FORTYTWO}\citep{CLL16} and \textsc{42++}\footnote{The \textsc{42++} is a PTA data analysis software package developed with the programming language C++ dedicated to the CPTA DR1 analysis. The software is available at \url{https://psr.pku.edu.cn/index.php/publications/gravitational-wave-data-analysis-code/}}. Consistent results were produced by the four pipelines (Chen et al., in prep.). We further compared all 16 possible combinations of noise modeling, which use/do not use EQUAD, ECORR, red noise, and DM noise. To avoid overfitting, the final best model is picked by comparing the Bayesian evidence of all 16 models. The data quality of the CPTA DR1 is summarized in Figure \ref{fig:data}. The detailed evaluation of how the noise modeling affects the GWB inference will be published by Chen et al. (in prep).

\section{Parameter inference for the stochastic GWB}
\label{sec:gwa}

We used the standard frequency-domain Bayesian method \citep{temponest,VV14} to perform statistical inference for the amplitude and spectral index of the stochastic GWB. We adopted the following definitions for the amplitude and spectral index of the GWB. The characteristic strain, $A(f)$, of GWB is
\begin{equation}
	A(f)=A_{\rm c} \left(\frac{f}{\rm 1 yr^{-1}}\right)^{\alpha}\,,	
	\label{eq:chara}
\end{equation}
where $A_c$ is the characteristic amplitude at the frequency of yr$^{-1}$, $\alpha$ is the spectral index for characteristic amplitude. For a stochastic GWB induced by the GW-driven merger of SMBHB population, $\alpha=-2/3$ is expected when the number of GW sources in the frequency band is sufficiently larger than unity \citep{Phinney01}.  The corresponding single-sided spectral density, $S(f)$, for the GW-induced pulsar timing residuals, i.e. differences between observed TOAs and the model predictions, is \citep{JHLM05}
\begin{equation}
	S(f)=\frac{A(f)^2}{12 \pi^2 f^3}, \,{\rm for }\, f>0\,.
	\label{eq:sf}
\end{equation}
With above definitions, the root-mean-square deviation (RMS) $\sigma$ of the GW-induced pulsar timing residuals is found by applying the Wiener–Khinchin theorem, as
\begin{equation}
	\sigma=\sqrt{\int_{f_{\rm low}}^{f_{\rm high}} S(f)\,df}\, ,
	\label{eq:rms}
\end{equation}
where $f_{\rm low}$ and $f_{\rm high}$ are the low and high boundaries defining the frequency band of the GW signal. 

We have used the parallel tempering Markov chain Monte Carlo (firstly applied in 
the PTA community by \citet{ptmcmc}) to perform the posterior sampling, where 
ten temperatures in a geometric series are used to speed up the initial burning 
runs and help to find the global maximum of the likelihood function (see the 
arguments of \citet{neal1996sampling} and \citet{VFM16}). The priors of the GWB 
parameters are uniform $\log A_{\rm c} \in [-18,-13]$ and $\alpha \in[-1.8, 
1.5]$. The prior range of $\alpha$ is determined by the requirement that the RMS 
of the stochastic power-law noise does not diverge after fitting the pulsar 
period and period derivative \citep{LBJ12}; the lower boundary of $-1.8$ is set 
slightly higher than the theoretical requirement, i.e. $\alpha>-2.0$, to avoid 
numerical singularities.

\begin{figure} \centering
	 \includegraphics[width=16.0cm, angle=0]{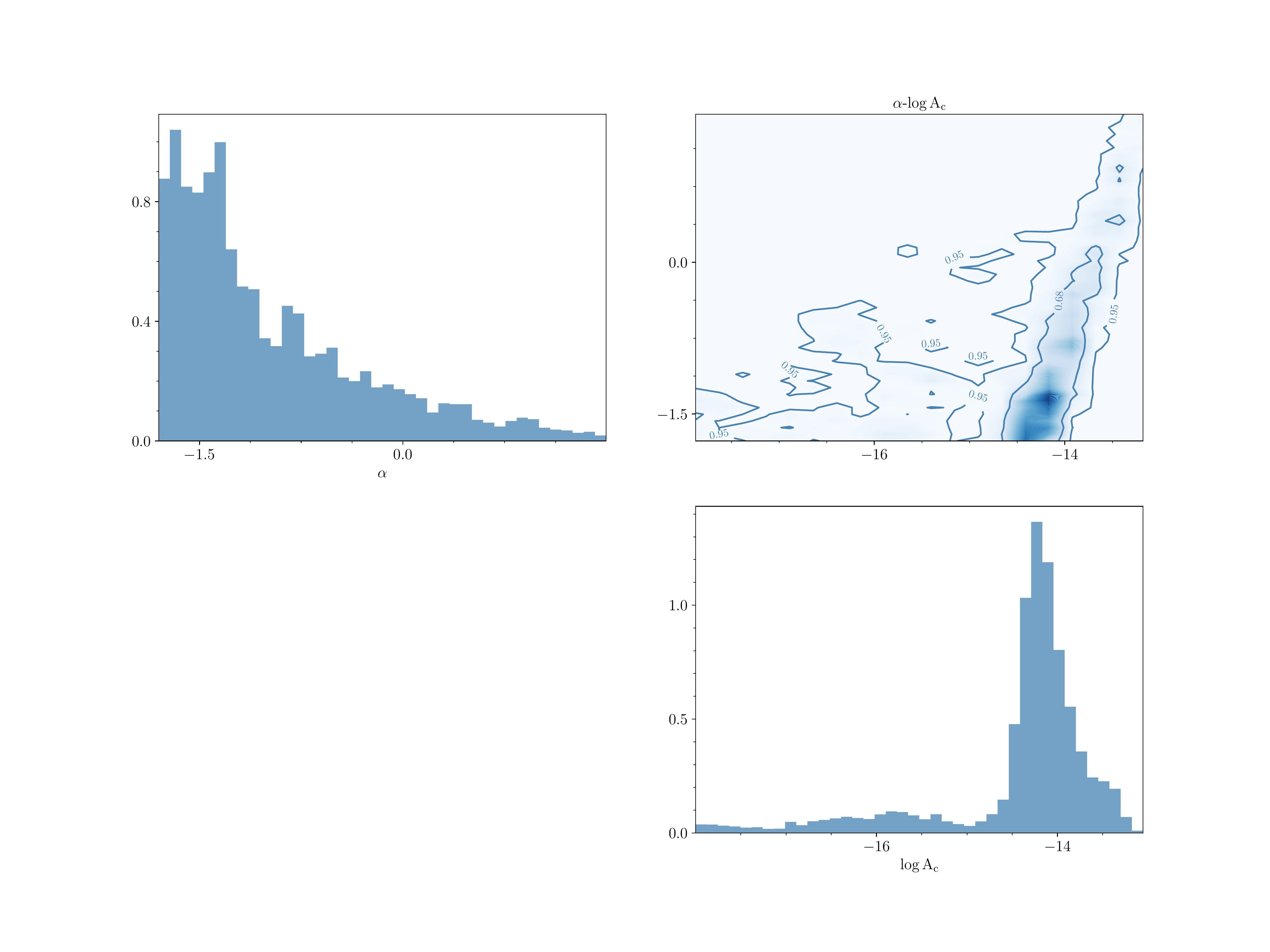}
	 \caption{The parameter inference for the stochastic GWB using CPTA DR1.  
	 Upper-left: histogram for the posterior distribution of spectral index 
	 $\alpha$. Upper-right: 2D distribution for the spectral index $\alpha$ and GW 
	 characteristic amplitude $A_{\rm c}$. Bottom-right: histogram for the 
	 distribution of GW characteristic amplitude.  \label{fig:gwa}}
   \end{figure}

\begin{figure} \centering
	\includegraphics[width=12.0cm, angle=0]{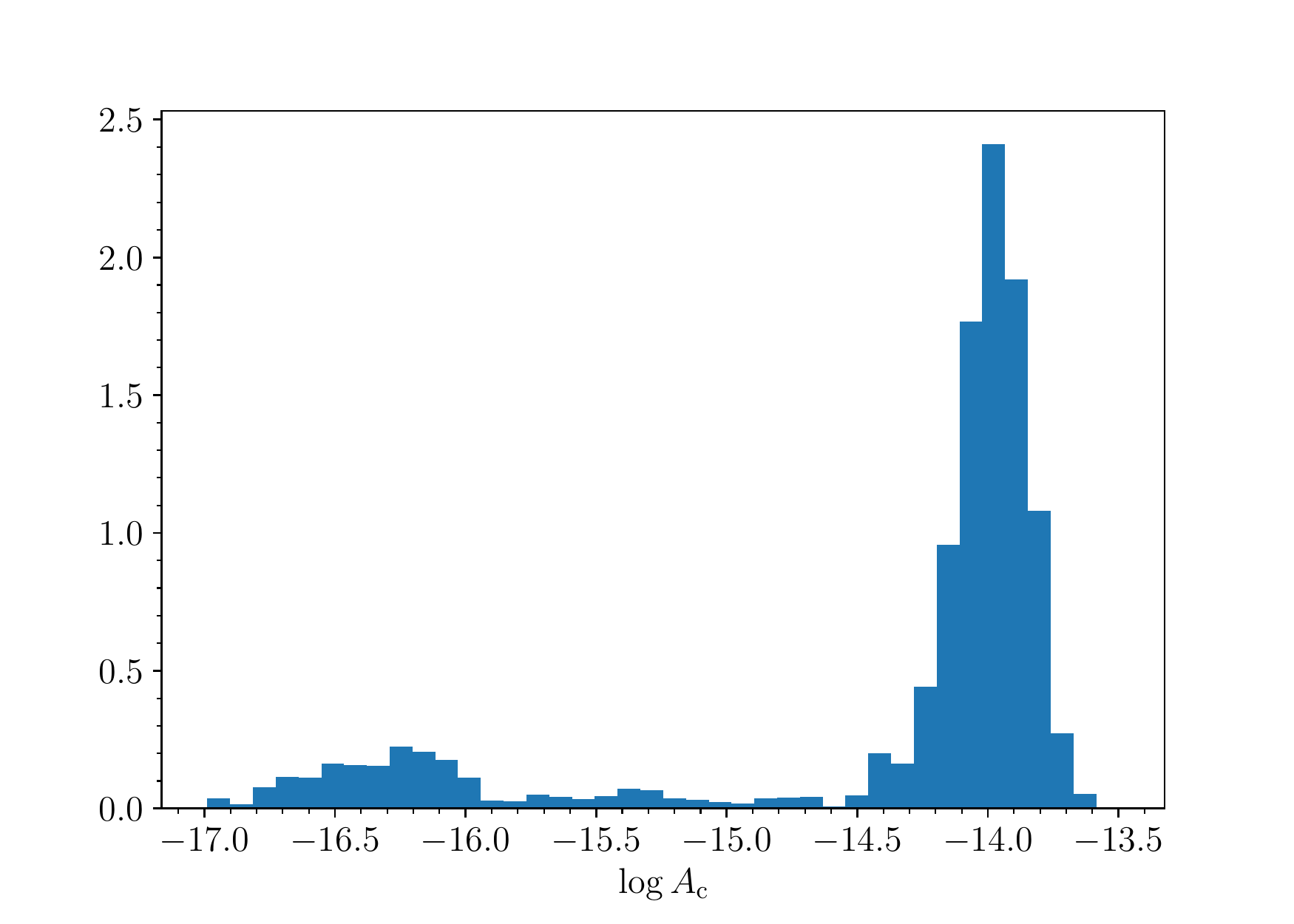}
	\caption{The histogram for the posterior distribution for the GWB amplitude after fixing the spectral index at $\alpha=-2/3$.  
	 \label{fig:gw23}}
\end{figure}

Our posterior distribution for the GWB parameters after marginalizing the pulsar timing and noise model is shown in \FIG{fig:gwa}, where we marginalized both pulsar timing parameters and noise parameters. Note that we have marginalized over all possible noise parameters without fixing any of them, i.e. our free parameters included white noise parameters, i.e., EFAC, EQUAD, and ECORR, as well as the red and DM noise amplitudes and spectral indices. Using the maximum likelihood estimator for the stochastic GWB amplitude, we recover $\log A_{\rm c}= -14.4 \,^{+1.0}_{-2.8}$ (for 95\% confidence level), while, because of the limited data span of only about 3 years, the spectral index $\alpha$ is not well constrained given our prior range of $-1.8$ to $1.5$. Even so, the distribution of $\alpha$ indicates that the signal is stronger at lower frequencies. If we fix the spectral index to be $\alpha=-2/3$ as predicted by the model of GW-driven merger of SMBHBs, the GWB amplitude is $\log A_{\rm c}=-14.7 \,^{+0.9}_{-1.9}$ (for 95\% confidence level) and the posterior is shown in \FIG{fig:gw23}.

\section{Searching for the Hellings-Downs curve}
\label{sec:hd}

PTAs search for the HD correlation signature \citep{HD83} to verify that the correlations in pulsar timing residuals are quadrupolar and induced by an isotropic stochastic GWB. We have searched for the HD correlation in the CPTA DR1. As explained in the previous section, the spectral index of the signal was not well constrained, because of the limited data span. Therefore, instead of searching for the spatial correlation in the framework of power-law spectral models, we performed a more robust search for the spatial correlation at discrete frequencies, such that 1) the results do not depend on the priors of how the correlation function is parameterized and 2) the results do not depend on the spectral shape information. We searched mainly at $f_{1.5}=14.0$ nHz, which corresponds to $1.5/T$ with $T=3.40$~years being the total time span of the CPTA DR1. The reasons for choosing such a frequency are explained in \APP{app:erf}. To check the consistency of results, we also searched at the other two frequencies, i.e., $f_1=9.34$ and $f_2=18.7$ nHz, respectively. These frequencies correspond to $1/T$ and $2/T$.

A two-step method was used to measure the pulsar-pair correlation coefficients for each discrete frequency.  We simultaneously fitted all pulsar timing residuals with the individual pulsar timing models and a common sinusoidal waveform at a given frequency $f$. After marginalizing the pulsar timing models, the phases of sinusoidal waveforms for each pulsar and the common amplitude were measured using the maximum likelihood estimator as
\begin{equation}
	A,\phi_i={\rm argmax}_{A,\phi_i}\int\int\dots\int\frac{1}{\sqrt{\prod_i 
	|C_i|}} \exp \left[ -\frac{1}{2}\sum_i \VEC{r}_i'^{T} C_i^{-1} \VEC{r}_i' 
	\right]\prod_{i} d \VEC{\lambda}_{T,i}\,,  \label{eq:phiaesti}
\end{equation}
were $\VEC{r}_i'$ is
\begin{equation}
	\VEC{r}_i'=\VEC{r}_i -\VEC{D}_i \VEC{\lambda}_{T,i}-A \sin (2 \pi f 
	\VEC{t}-\phi_i)\,.
	\label{eq:reshd}
\end{equation}
Here, the subscript $i$ denotes the index of pulsar. Vector $\VEC{r}_{i}$ and matrix $\VEC{D}_{i}$ are the timing residual and timing model design matrix for the $i$-th pulsar, respectively. $\VEC{\lambda}_{T,i}$ are pulsar timing and noise parameters, e.g. pulsar period, period derivative, white, red and DM noise. $C$ is the pulsar noise covariance matrix. $\phi_i$ is the phase of the correlated signal for the $i$-th pulsar. $A$ is the amplitude of the correlated signal in timing residual, e.g. a GWB-induced signal. Here, amplitude $A$ is proportional to the RMS defined in \EQ{eq:rms} over a narrow frequency band around frequency $f$. However, due to the irregular sampling of pulsar timing data, there is no close formula to express $A$ in terms of spectral density $S(f)$.  One can show that this approach is equivalent to the frequency-domain modeling of power-law processes \citep{VV14} with one frequency element. The pairwise correlation coefficient between the $i$-th and the $j$-th pulsars is $c_{ij}=\cos(\phi_i-\phi_j)$. 

The statistical significance for the HD correlation against constant correlation is evaluated using the frequentist method described by \citet{JHLM05}, as
\begin{equation}
	{\cal S}\equiv\sqrt{\frac{N (N-1)}{2}} \frac{\sum_{i<j} (c_{ij} - \overline{c_{ij}}) 
	(H_{ij} -\overline{H_{ij}})}{\sqrt{\sum_{i<j} (c_{ij} - \overline{c_{ij}})^2 
	\sum_{i<j} (H_{ij} -\overline{H_{ij}})^2  }}\,, \label{eq:decs}
\end{equation}
with the average operator over pulsar pairs defined as $\overline{x_{ij} }\equiv \frac{2}{N(N-1)} \sum_{i<j} x_{ij}$, where $N$ is the number of pulsars, $\sum_{i<j}$ sums over all independent pulsar pairs, and $H_{ij}$ (for $i\neq j$) is the HD function of pulsar pairs that
\begin{equation}
H_{ij}=\frac{1}{2}-\frac{1}{4} \left(\frac{1-\cos \theta_{ij}}{2}\right)+\frac{3}{2} \left(\frac{1-\cos \theta_{ij}}{2}\right)\log \left(\frac{1-\cos \theta_{ij}}{2}\right)\,,
\end{equation}
where $\theta_{ij}$ is the angular distance between the $i$-th and the $j$-th pulsars.
The exact expression for the null-hypothesis distribution of $\cal S$ is rather lengthy \citep{ks39}. For our case, where the number of pulsars is significantly larger than unity, a much simpler asymptotic Gaussian form is available as explained by \citet{JHLM05}, i.e., the probability density distribution function of $\cal S$ with no correlation is
\begin{equation}
    f({\cal S})=\frac{1}{\sqrt{2\pi} }\exp\left(-\frac{{\cal S}^2}{2} \right)\,.
    \label{eq:sdis}
\end{equation}
It should be noted that this method is immune to the contamination of common uncorrelated signals \citep{EPTA15,NANOGRAV,EPTA21,2022MNRAS.510.4873A,2021ApJ...917L..19G}, because it evaluates the statistical significance only using the cross correlation. Furthermore, the monopolar correlation induced by clock errors is also automatically removed, because the average value of the correlation coefficients is subtracted in \EQ{eq:decs}. Another interesting property of the above statistic is that the error in $c_{ij}$ is regularized such that $-1\le c_{ij}\le 1$. On the one hand, this makes the $\cal S$ not the optimal statistic to search for the HD curve. On the other hand, the error of $c_{ij}$ becomes weakly dependent on the pulsar intrinsic noise, and $\cal S$ is less affected  by the systematics of a few pulsars with dominant precision. We demonstrate the application of this method with two simulated datasets, where one contains the GWB signal injection (positive control group) and the other does not (negative control group). The simulated datasets are the `clones' of the CPTA DR1 that they have the same frequency resolution and sampling epochs as the CPTA DR1.

\begin{figure} \centering
	 \includegraphics[width=15.0cm, angle=0]{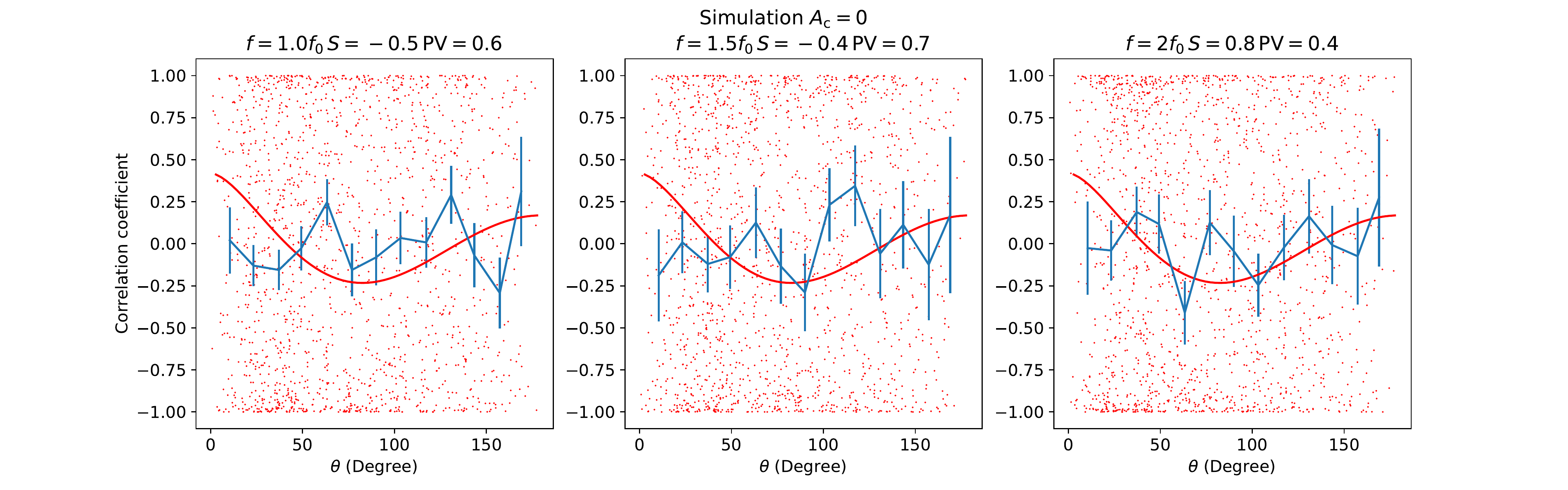}
	 \includegraphics[width=15.0cm, angle=0]{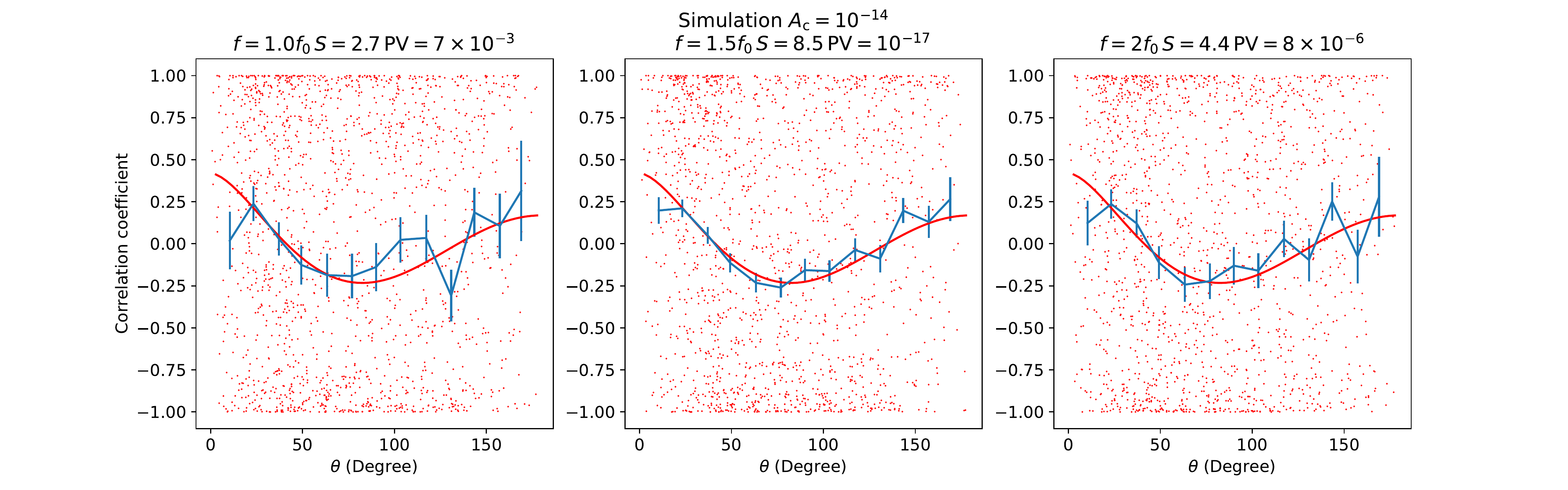}
	 \includegraphics[width=15.0cm, angle=0]{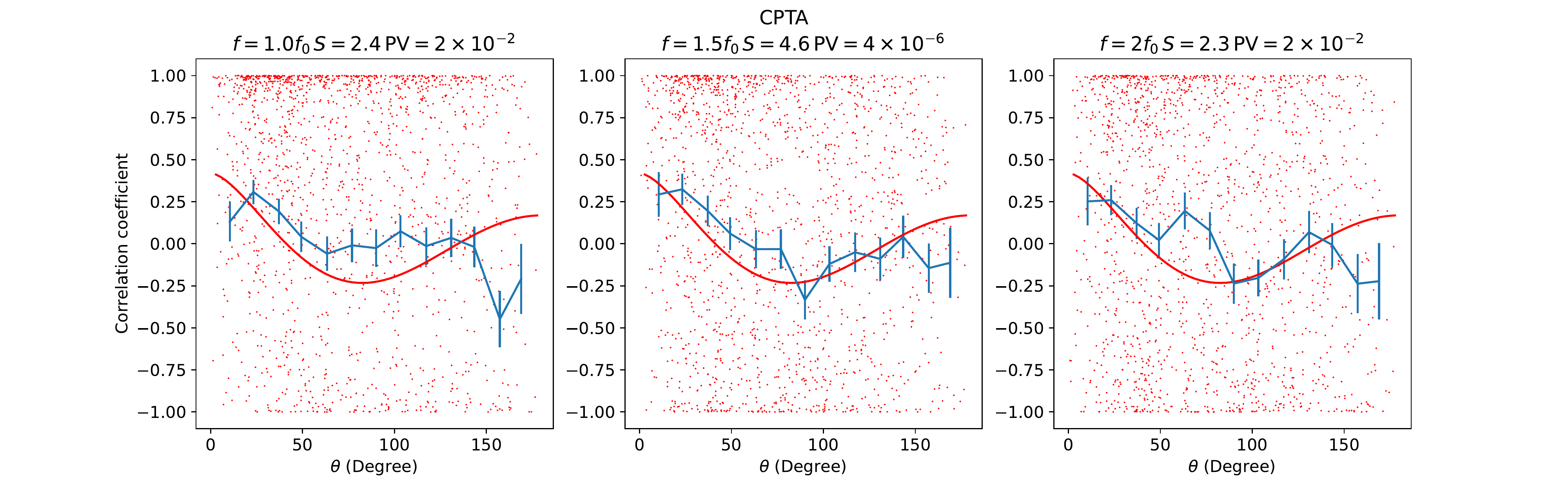}
	 \caption{The measured correlation coefficients (y-axis) as a function of the pulsar-pair 
	 separation angle (x-axis). Red dots denote the measured correlation 
	 coefficients between all pulsar-pairs without the auto-correlation.  The blue 
	 curves with error bars represent the binned average red dots, which only serve to aid the visual 
	 inspection. The error bars are the standard error of binned average value estimated using the binned red dots. The solid red curves depict the theoretical HD curve. The top row three 
	 panels show simulations without the GWB signal injection, where the data was simulated to match exactly the times and frequencies of the real CPTA DR1.  Each panel from left to right corresponds to $f=1/T, 1.5/T$, and $2/T$, respectively.  
	 \label{fig:hdfun}}
   \end{figure}

The measured pair-wise correlation coefficients of CPTA DR1 are shown in \FIG{fig:hdfun}, where the results of the negative and positive control groups using simulated datesets are also shown. For the simulated dataset without the GWB injection (i.e. the negative control), we detect no significant spatial correlations as shown in the top panels of \FIG{fig:hdfun}.

We can detect the HD correlation in the positive control with the injected GW 
signal as shown in the middle panels of \FIG{fig:hdfun}. In the dataset, the 
characteristic amplitude and spectral index of GWB are $A_{\rm c}=10^{-14}$ and 
$\alpha=-2/3$, respectively.  The statistical significance (${\cal S}=8.5$) 
peaks at $f=1.5/T$, whereas it is low at $f_1$ and $f_2$ (${\cal S}=2.7$ and 
${\cal S}=4.4$, respectively). Such behaviour agrees with the theoretical 
expectation, as explained in \APP{app:erf}, that 1) the statistic significance 
of $f=1/T$ is lower than that at $f=1.5/T$, because fitting of the pulsar timing 
model affects the low-frequency components of the signal; and 2) the statistic 
significance at $f=2/T$ is lower than that at $f=1.5/T$, because the power-law 
spectral signal is weaker at higher frequency.

The statistical significance of the HD correlation of CPTA DR1 shows similar features to those of the positive control group. For CPTA DR1, as shown in the bottom panels of \FIG{fig:hdfun}, ${\cal S}=4.6$ peaks at $1.5/T$, which corresponds to a P-value of $\rm PV=4\times 10^{-6}$.
As we saw in the positive control group, ${\cal S}=2.4$ and $2.3$ are lower at $f=1/T$ and $2/T$, respectively. By comparing the positive control group results and the CPTA DR1 results, we can further confirm that the GWB amplitude should be lower than $10^{-14}$. We also performed the phase shift and sky position scrambling experiments, which are described in \APP{app:scram}. 

The signal components of the three frequencies $1/T$, $1.5/T$, and $2/T$ are not independent of one another, because the sampling of the data is irregular. It is thus invalid to directly add (in the sense of the square-root sum of squares) the statistical significance at the three frequencies to compute the frequency-integrated statistical significance, even though the frequency-integrated statistical significance is always higher or equal to the statistical significance measured at a single discrete frequency. 
In the case where noise modeling is not accurate enough, the frequency-integrated statistical significance could be lower than the single frequency value because of statistical fluctuations or wrong weights at some frequencies. This could happen if longer data is used and the signal spectrum over a wide frequency range can not be well described by a single power law.  

For the CPTA DR1, we can not discriminate between dipole (cosine function) and HD correlation with only cross-correlation coefficients, although it seems to be lack of physical mechanism to produce the dipole correlation at the frequency we are currently sensitive to. For dipole correlation, we get ${\cal S}=4.1$ at $f=1.5/T$, which is only slightly lower than that of HD correlation. If we use the Bayes factor as the statistic (see the method described in \citet{NANOGRAV}), HD correlation is preferred with a `strong evidence' that the Bayes factor ${\cal B}|_{\rm HD/dipole}=66$. 
We should point out that one of the major differences between the Bayesian method and the above frequentist method is that the Bayesian method includes the autocorrelation terms in comparing the two models, while the above method does not. Furthermore, Bayes factor ${\cal B}|_{\rm HD/dipole}=66$ here is not the perfect touchstone to exclude the dipole correlation interpretation, as demonstrated by numerical simulations of \citet{redherr} and the toy model in \APP{app:bf}. One needs to be cautious about the application and interpretation of the Bayes factors for the current problem of measuring or detecting the \emph{statistical variance} of stochastic signals with spatial correlations.

\section{Discussion and conclusion}
\label{sec:con}

In this paper, we show that the inferred GWB characteristic amplitude is $\log 
A_{\rm c}= -14.4 \,^{+1.0}_{-2.8}$ for a spectral index in the range $\alpha\in 
[-1.8, 1.5]$, and $\log A_{\rm c}=-14.7 \,^{+0.9}_{-1.9}$ if fixing 
$\alpha=-2/3$. The measured GWB amplitude agrees with theoretical expectation 
\citep{SESANA13,2014ApJ...789..156M}.  However, because of the short span of the 
CPTA DR1, we cannot yet differentiate between different models of SMBHB 
formation and evolution. Our GWB amplitude seems to be compatible with that of 
the common red signal of other PTAs 
\citep{EPTA21,NANOGRAV,2022MNRAS.510.4873A,2021ApJ...917L..19G}, although the 
median of posterior distribution seems to be 0.4 to 0.5 dex higher. It is 
possible that a single power law with $\alpha=-2/3$ is insufficient to extend 
the previous results to the high frequency band that CPTA is currently sensitive 
to, i.e., the spectrum of the GWB signal is not fully power-law in shape and may 
have a spectral `bump' in the CPTA frequency band ($\ge 10^{-8}$ Hz). The 
statistical significance of the HD correlation over a constant-value correlation 
in CPTA DR1 is 4.6-$\sigma$ around 14 nHz, i.e., a P-value of $4\times 10^{-6}$. 
Our correlation-curve analysis is compatible with the GWB amplitude inference, 
where the comparison between the statistical significance of simulations and  
CPTA DR1 suggests that the amplitude of GWB quadruple signal $A<10^{-14}$.  More 
details, including comparisons between different data reduction pipelines and a 
cross-check of the current measurements, will be published in a following paper 
(Xu et al., in prep.). 

For the spatial correlation inference, we measured the  pulsar pairwise correlation at single frequencies, which removes the power-law presumption for the GWB spectral shape. Additionally, this method is not affected by common uncorrelated noise or clock error, as it uses only the cross correlations and subtracts their average value. However, this method limits our statistical significance, because only the correlation at a single frequency was extracted. The total statistical significance will be higher than the single frequency values. However, to combine the measurements at multiple frequencies to deliver the frequency-integrated spatial correlation, we would need accurate information on the GWB spectral shape and pulsar noise properties. In the future, we expect that data with a longer span will enable us to go lower in frequency and therefore measure the spectral index with better accuracy. 

The current method cannot rule out a dipole origin for the correlation, since both dipole and HD correlations produce similar values of $\cal S$. If we use Bayesian method to compare the models of dipole and HD correlation, the Bayes factor (with the caveats discussed in \APP{app:bf}) prefers the HD correlation that the Bayes factor of HD correlation over dipole correlation is ${\cal B}|_{\rm HD/dipole}=66$. 

The current CPTA DR1 statistical significance is still below the IPTA `detection' bar of ${\cal S}=5$. Independent results of other regional PTAs may soon help to confirm the current findings. On a longer timescale, we look forward to officially joining the IPTA. By combining IPTA and CPTA data, we expect a further increase in GWB detection sensitivity. The CPTA DR2, scheduled for 2026, is also expected to deliver better accuracy in GWB parameter inference. We are clearly in the era of opening the nHz GW observation window.

\normalem

\begin{acknowledgements}

Observation of CPTA is supported by the FAST Key project. FAST is a Chinese national mega-science facility, operated by National Astronomical Observatories, Chinese Academy of Sciences. This work is supported by the National SKA Program of China (2020SKA0120100), the National Nature Science Foundation grant no. 12041303 and 12250410246, the CAS-MPG LEGACY project, and funding from the Max-Planck Partner Group. KJL acknowledges support from the XPLORER PRIZE and 20-year long-term support from Dr. Guojun Qiao. The data analysis are performed with computer clusters \textsc{DIRAC} and \textsc{C*-system} of PSR@pku and computational resource provide by the \textsc{PARATERA} company. Software package 42++ is developed with \textsc{Intel} oneAPI toolkits and the \textsc{Science Explorer's Developing GEars} from \textsc{Weichuan technology}. We thank Dr. Jin Chang for providing the valuable long-term support to the CPTA collaboration. CPTA thanks Dr Duncan Lorimer and Dr Michael Kramer for helping initiate the collaboration. 

HX, SYC, YJG, JJC, BJW, JWX, ZHX, RNC, JPY, YHX, JBW and KJL are core team to perform the data analysis for the current paper, where HX worked on data reduction, timing and data analysis; HX, SYC, YJG and KJL performed the noise and GWB data analysis; BJW and JPY carried out single pulse studies; JJC and JWX calibrated data polarization; ZHX and RNC searched for GW single sources; YHX studied scintillation processes; and JBW studies the GW memory effects. Before FAST observation, LFH and JTL helped test pulsar timing observation with Luonan and Kunming 40m radio telescopes.  KJL organized the team and drafted the current paper. In alphabet order, JLH, PJ, ZQS, MW, NW, and RXX  are the CPTA executive committee members. XPW and RM served as the supervise committee members oversee the CPTA collaboration. RM helped create the pulsar community in China and the formation of the CPTA. QL, XG, MLH, CS, and YZ are the core FAST support team members.

\end{acknowledgements}


\appendix
\section{Is the Bayes factor invalid? - analytic study of a toy model}
\label{app:bf}
Here we provide a toy model to illustrate the problems of applying the Bayes factor in the context of \emph{common signals}.
Our toy model contains $N$ pulsars. The timing residuals of all pulsars are just pure white noise. There is neither spatially correlated nor common uncorrelated signal components. The null and positive hypotheses are
\begin{equation}
	\left\{ \begin{array}{l}
		H_0\textrm{: pulsar timing residuals are described by intrinsic noise only;} \\
		H_1\textrm{: intrinsic noise and a common uncorrelated component are needed.}
	\end{array}\right.
\end{equation} 
We further simplify the toy model by setting the average value of timing residual to be 0. For $H_0$, the parameters of the noise model contain the standard deviation of the timing residual, which is denoted as $\sigma_i$ for the $i$-th pulsar. 
The likelihood of $H_0$ is
\begin{eqnarray}
	\Lambda_0(\sigma_1, \sigma_2\ldots\sigma_N)\propto\prod_{i=1}^{N}\frac{1}{\sigma_i^{N_{pt, i}}}
	\exp\left[-\frac{1}{2} \frac{\VEC{r_i} \cdot \VEC{r_i}}{\sigma^2_{i}} \right]
\end{eqnarray} 
The expected value of Bayes evidence (${\rm BE}$) with logarithmic prior (the most common choice in PTA problems) is
\begin{equation}
	\langle{\rm BE_0}\rangle = \int\ldots\int \Lambda_0(\sigma_1, \sigma_2\ldots\sigma_N) d \log \sigma_1\,\ldots d \log \sigma_N\propto\prod_{i=1}^{N} 2^{\frac{N_{pt,i}-2}{2}} (N_{pt,i} \sigma_i^2)^{-\frac{N_{pt,i}}{2}} \Gamma\left(\frac{N_{pt,i}}{2}\right)\,.
\end{equation}
For $H_1$, an extra amplitude parameter ($A$) is required to model the amplitude of the common uncorrelated signal, so that the likelihood function is
\begin{eqnarray}
	\Lambda_1(\sigma_1, \sigma_2\ldots\sigma_N; A)\propto\prod_{i=1}^{N}\frac{1}{(\sigma_i^2+A^2)^{N_{pt, i}/2}}
	\exp\left[-\frac{1}{2} \frac{\VEC{r_i} \cdot \VEC{r_i}}{\sigma^2_{i}+A^2} \right]\,.
\end{eqnarray} 
Clearly, the expectation of Bayes evidence of $H_1$, i.e.
\begin{equation}
	\langle{\rm BE_1}\rangle=\int\ldots\int \Lambda_1(\sigma_1, \sigma_2\ldots\sigma_N) d \log \sigma_1\,\ldots d \log \sigma_N\,d A\,,
\end{equation}
will not converge on the boundary of $\partial|\sigma_i=0$, as the amplitude $A$ modifies the behavior of the exponential function when $\sigma_i\to0$ and the singularity rises because of the logarithmic prior. In other words, the Bayes factor $\rm BE_1/BE_0$ can be arbitrarily large no matter what the data is, if one varies the prior range in the current toy model. This is true even the prior range is kept to be the same for both $H_0$ and $H_1$! It is not hard to see that the similar argument is also applicable to the case of comparing two spatially correlated signals, where one will be evaluating the Bayes factor between two singular distributions. Numerical simulations have shown similar features, we refer interested readers to the work of \citet{redherr}.

Furthermore, although one can regularize the prior singularity by using finite priors, the Bayes factor then becomes prior dependent. Other singular behaviors will rise, when the spatial correlations and spectral properties of signals are considered. All those effects indicate that we should be cautious about applying and interpreting the Bayes factors in the PTA GWB searching problems. To fully utilize the power of Bayes factors, we need 1) the probability distribution function of Bayes factors under the null hypothesis, which requires that the sample space is measurable such that probability distribution function can be well defined, and 2) the computational method to calculate the null-hypothesis distribution of Bayes factors that converges to the full-sample-space expectation at the large N limit.

\section{Systematic error of correlation coefficients due to the fitting of 
pulsar period and period derivative}

\label{app:erf}
For two cosine functions with the same frequency, i.e. $f_1=A_1\cos(2\pi f t+\phi_1)$ and $f_2=A_2\cos(2\pi f t+\phi_2)$, the correlation coefficient is $c=\cos(\phi_1-\phi_2)$. Here one can regard the two cosine functions as the single-frequency components of GW-induced signals for two different pulsars. In the pulsar timing procedure, one fits for the pulsar period and period derivative. Thus, the best-fitted quadratic is subtracted by-default from the timing residuals.
Such a fitting modifies the correlation coefficient between the two cosine functions and leads to a systematic error.   

After subtracting the best-fitted quadratic, the two functions are  $f'_1=A_1\cos(2\pi f t+\phi_1)-\alpha_0 -\alpha_1 t -\alpha_2 t^2$ and $f'_2=A_2 \cos(2\pi f t+\phi_2)-\beta_0 -\beta_1 t- \beta_2 t^2$, where the coefficients of the quadratic, i.e. $\alpha_{0..2}$ and $\beta_{0..2}$, are found by minimizing the $\chi^2$, i.e.
\begin{eqnarray}
	\alpha_{0..2}={\rm argmin}_{\alpha_{0..2}} \int_0^{T} \left[A_1\cos(2\pi f t+\phi_1)-\alpha_0 -\alpha_1 t -\alpha_2 t^2\right]^2 \,dT\,. \\
 \beta_{0..2}={\rm argmin}_{\beta_{0..2}} \int_0^{T} \left[A_2\cos(2\pi f t+\phi_2)-\beta_0 -\beta_1 t -\beta_2 t^2\right]^2 \,dT\,.
\end{eqnarray}

\begin{figure} \centering
	\includegraphics[width=10.0cm, angle=0]{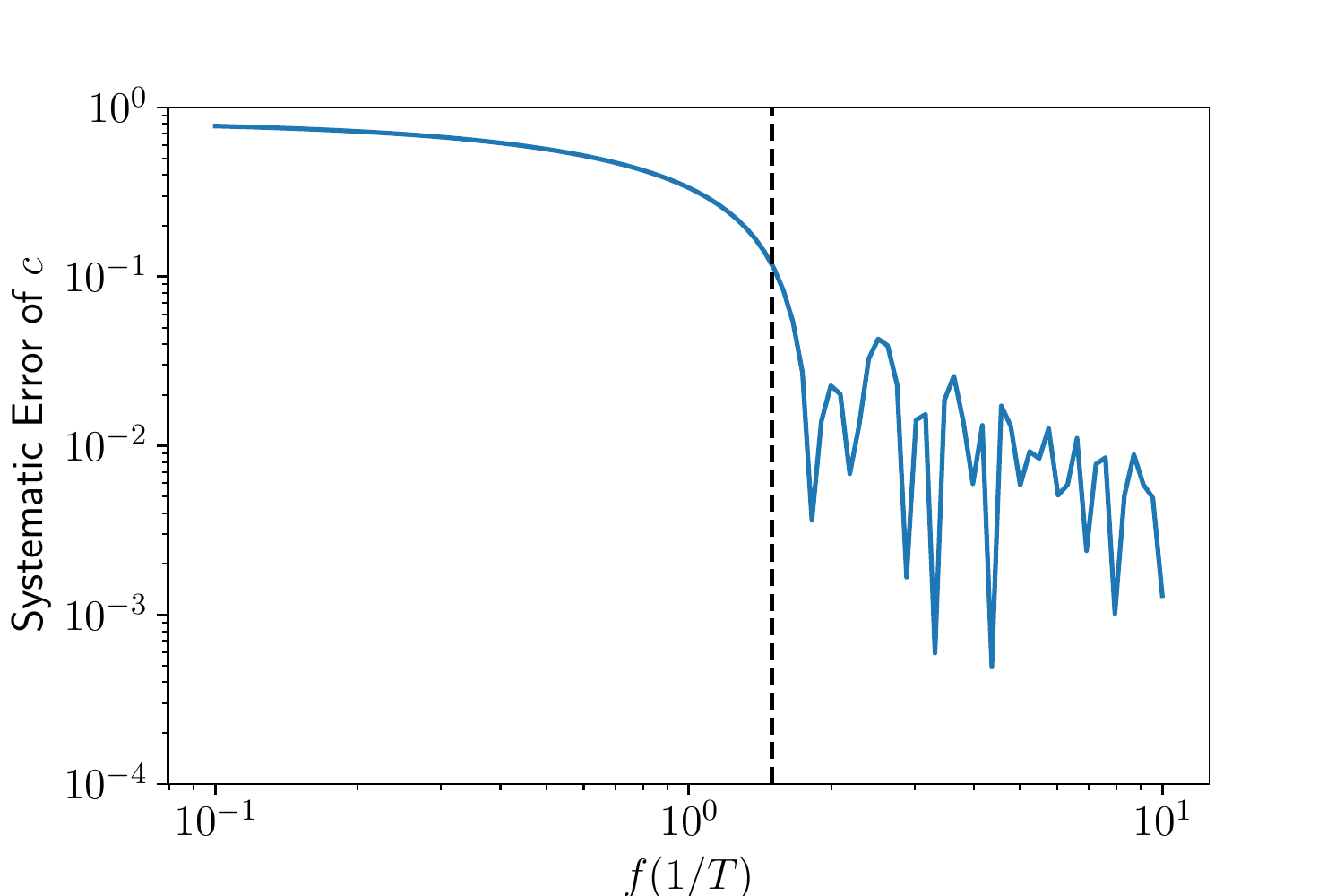}
	\caption{The systematic error of correlation coefficients $c_{ij}$ due to the quadratic fitting. Here, x-axis is the frequency in the unit of $1/T$, and y-axis is the systematic error of correlation coefficients defined in \EQ{eq:cerr}. The vertical dashed line indicates $f=1.5/T$.  
		\label{fig:cijerr}}
\end{figure}

To evaluate the effects of the above quadratic fitting on the correlation coefficient $c$, we define the RMS level of systematic error over all possible correlations as
\begin{equation}
	\delta c\equiv\sqrt{\left\{\frac{1}{4\pi^2}\int_0^{2\pi}\int_{0}^{2\pi} \left[\frac{\int_0^T f'_1 f'_2 dt }{\sqrt{\int_0^T f'_1 f'_1 dt \int_0^T f'_2 f'_2 dt} }-\cos(\phi_1-\phi_2)\right]^2 d\phi_1 d\phi_2 \right\}}\,.
	\label{eq:cerr}
\end{equation} 
The systematic error as a function of the frequency $f$ is shown in \FIG{fig:cijerr}. As expected, it shows that the correlation coefficients of lower frequency components ($f\le1/T$) are mostly affected. In the paper, we choose $f=1.5/T$, such that the systematic error of the correlation coefficient is less than 10\%. For future datasets, the frequency can be chosen with a higher value to further reduce the systematic error, as the data sensitivity improves.

\section{Phase shift and Sky-position scramble experiments}
\label{app:scram}
As is common in PTA data analysis at the time this paper is written, we perform 
the experiments with the phase shift and the sky-position scramble, which were 
firstly introduced by \citet{2017PhRvD..95d2002T}. The recipes for the two 
experiments are as follows: (i) For the phase shift experiment, one introduces 
random phase either to the data or Fourier design matrix, which eliminates the 
phase coherence between pulsars. With each phase shift, the desired statistics 
in a given GW detection pipeline are computed and collected. The distribution of 
the collected values of the statistics is then used to form an approximation to 
the null-hypothesis statistical distribution function to evaluate the false 
alarm probability. (ii) The sky-position scramble is similar, where one replaces 
the phase shifting with the scrambled pulsar position, i.e. assigns random 
positions to the pulsars. The method aims at shuffling the pulsar position to 
remove the spatial correlation, where one randomly re-assigns all pulsars with 
new pulsar positions over the entire sky and ensures that the `matching' (see 
definition in \citealt{2017PhRvD..95d2002T}) is below a prescribed threshold 
($M$) between the scrambled spatial correlation functions. Note, it is required 
that the matching between any two full skies of the position-scrambles is also 
below the threshold. 

\begin{figure}
\centering
\includegraphics[width=18.0cm, angle=0]{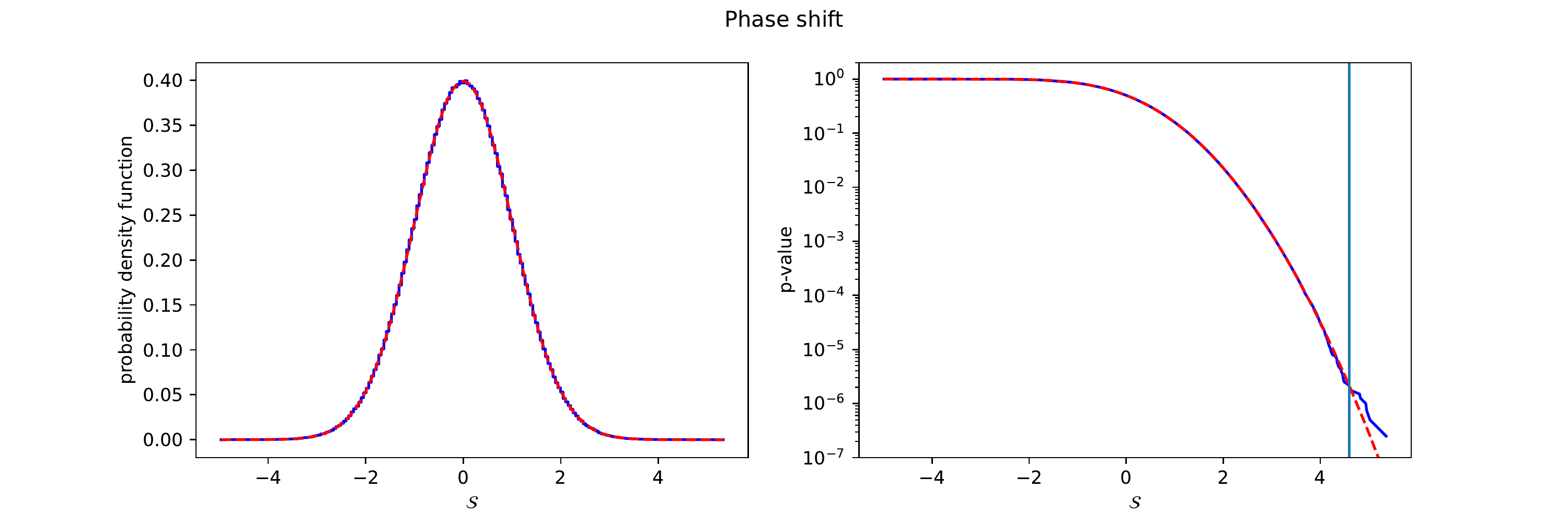}
 \includegraphics[width=18.0cm, angle=0]{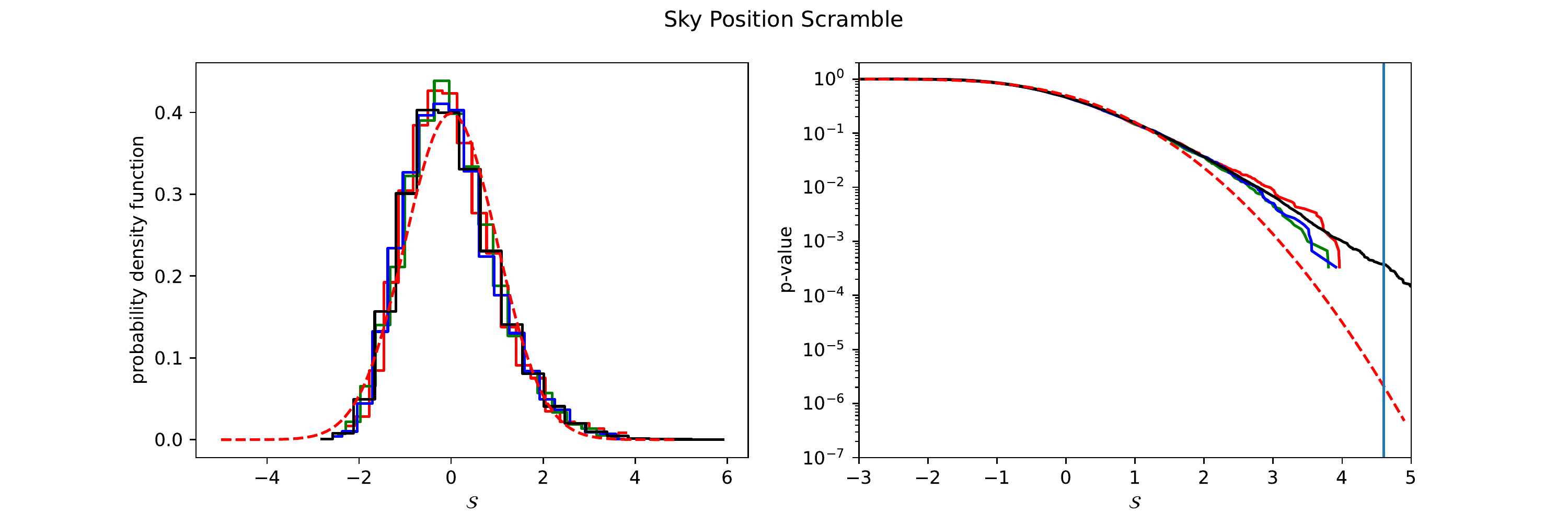}
	\caption{Phase shifting and sky position scramble experiments.  Upper row: The probability density function (upper left) and the survivor function (upper right) of $\cal S$, respectively, after introducing random phase shifts. The blue curves are measured by evaluating $\cal S$ with each realization of phase shift. The red dashed curve are the Gaussian distribution (\EQ{eq:sdis}). The vertical blue line indicates ${\cal S}=4.6$.
 Lower row: The probability density function (lower left) and the survivor function (lower right) of $\cal S$, respectively, after the sky position scrambling. The red, green, blue, and black curves are for matching thresholds of $M=0.1$, 0.3, 0.5, and 1, respectively. Because of the matching-threshold's limitation in the number of possible realizations,  we create 3000 realizations for $M=0.1,0.3,$ and $0.5$, and 100000 realizations for $M=1$. The differences between those curves are dominated by the statistical fluctuation, and there is no significant difference with choosing different values of matching threshold. However, the distribution of ${\cal S}$ with sky scrambling differs significantly from the Gaussian distribution (the red dashed curve) when ${\cal S}\ge 2$, as seen clearly in the survivor functions (right panel). 
		\label{fig:scam}}
\end{figure}

For our statistics $\cal S$ defined in \EQ{eq:decs}, the results of phase shifting and sky position scrambling experiments are shown in \FIG{fig:scam}. One can see that the phase shifting reproduced the expected null hypothesis distribution. This is not surprising. For our statistics, as $\cal S$ does not depend on the amplitude of the correlated signal, the distribution of $\cal S$ becomes the null-hypothesis distribution after the correlation between any pulsar pair is destroyed by the random phase shifts. One can get different conclusions, if an alternative statistics involving the amplitude parameter is used, e.g. optimal statistics, likelihood ratio tests, and Bayes factors. 

\begin{figure}
\centering
\includegraphics[width=12.0cm, angle=0]{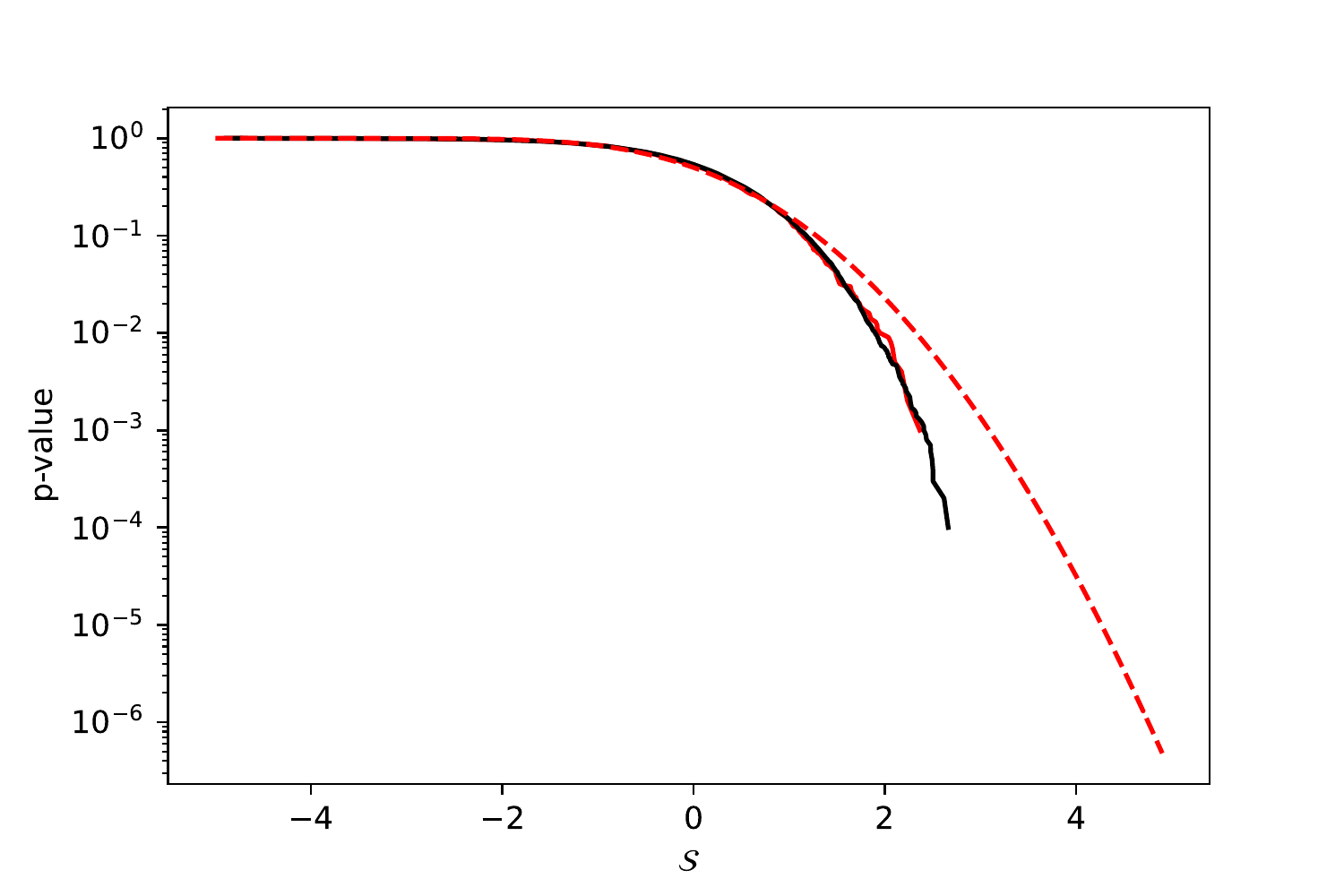}
	\caption{Sky position scramble experiments imposing shuffled anti-HD correlation.  For simplicity, we only show the survivor function of $\cal S$ here. The red and black solid curves are measured by evaluating $\cal S$ with each realization of phase shift with  $M=0.1$ and 1, respectively. The red dashed curve is for the Gaussian distribution \EQ{eq:sdis}. 
		\label{fig:nthel}}
\end{figure}

For sky-position scrambling, the distribution of $\cal S$ is similar to the null-hypothesis distribution for ${\cal S}\le 2$. However the position scrambling distribution has a significant tail towards higher value of $\cal S$, and sky scrambling \emph{overestimates} the p-value. The reason is that the sky scrambling produces a \emph{different sample space} comparing to the case of \EQ{eq:sdis}. The sample space produced by the sky scrambling still contains a shuffled version of the prescribed spatial correlation in the data set (the by-default choice is a shuffled HD correlation), while \EQ{eq:sdis} assumes that the sample space contains the data set with no spatial correlation.  We can demonstrate such a sample-space problem by computing the distribution of $\cal S$ while imposing another shuffled spatial correlation, which, according to the same argument by \citet{2017PhRvD..95d2002T}, still forms an uncorrelated distribution. For a sky scramble of an inverted HD correlation, the results can be found in \FIG{fig:nthel}. One can see that the sky scrambling now \emph{underestimates} the p-value. Clearly, all correlations do not die in the sky scrambling procedure.

\begin{figure}
\centering
\includegraphics[width=18.0cm, angle=0]{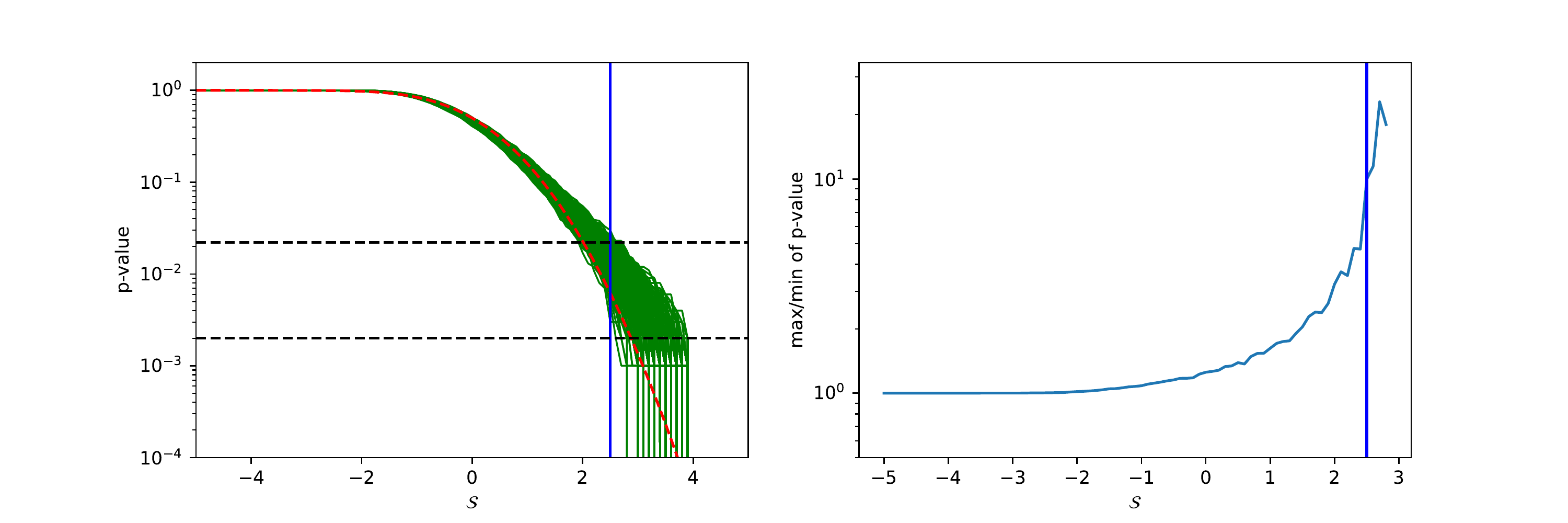}
	\caption{1000 $\times$ 1000 sky position scramble experiment imposing shuffled HD correlation. Left: The survivor function of $\cal S$. Each of the green curves is a single trial with 1000 realizations with $M=0.1$. The red dashed curve is for the Gaussian distribution \EQ{eq:sdis}. The blue vertical line indicates the ${\cal S}=2.5$, while the two black horizontal dashed lines indicate the maximal and minimal p-value at ${\cal S}=2.5$ found from the green curves, which are $2 \times 10^{-2}$ and $2 \times 10^{-3}$, respectively. Right: The scale of p-value fluctuation as a function of $\cal S$, computed from the green curves ensembles shown in the left panel. Here, we use the ratio between the maximal and minimal p-value to indicate the fluctuation scale. For ${\cal S} \ge 2.5$ (blue vertical line), the fluctuation is larger than a factor of 10. 
    In this way, a single trial with 1000 realizations is insufficient to accurately determine the p-value for ${\cal S} \ge 2.5$. Significantly more realizations are thus required.
		\label{fig:skfall}}
\end{figure}

In addition to the sample space issue, we note that the sky position scrambling operation has two more practical problems and should not be used to evaluate the p-value for our case of ${\cal S}=4.6$. The first issue is that as the original method 
\citet{2017PhRvD..95d2002T} requires that all realizations are below a certain matching threshold,
it is computationally expensive to generate a large number (e.g. $\sim10^{4}$) 
of samples; if one further requires the noise-weighted threshold (as in general 
not all pulsars contribute equally due to the differences in their noise 
properties), the number of independent samples can be further limited \citep{2023arXiv230504464D}.  
The second issue is that the distribution of $\cal S$ suffers from fluctuation 
of individual realizations. We would need thousands of samples to evaluate the 
statistical threshold for ${\cal S}\ge 3$. As an example, in \FIG{fig:skfall}, 
we show the survivor function of $S$ from 1000 tries each with 1000 
realizations. One can see that the p-value fluctuates from $2 \times 10^{-3}$ to 
$2 \times 10^{-2}$ at ${\cal S}=2.5$. Each realization in general over estimate 
the p-value compared to \EQ{eq:sdis}, although a few realizations underestimate 
the p-value. More than one order of magnitude fluctuation in p-value is noted, 
for ${\cal S} \ge 2.5$. The fluctuation can be even larger, once we allow for a 
more general form of imposed spatial correlation as explained in the previous 
paragraph. 

One may attempt to remedy the fluctuation by combining the phase shifting and sky position scrambling to produce a larger data set. This will not work in our case, as phase shifting and sky-position scrambling cover different sample spaces. The final p-value calculation depends on the details of how the two methods are combined, which is not meaningful in the statistical modeling. 

Due to the results of above experiments and reasons explained, we do not use phase shift or sky-position scrambling to evaluate the p-value of $\cal S$ in the current paper. 

\end{document}